\documentclass[conference]{IEEEtran}
\IEEEoverridecommandlockouts
\usepackage{cite}
\usepackage{amsmath,amssymb,amsfonts}
\usepackage{algorithmic}
\usepackage{graphicx}
\usepackage{textcomp}
\usepackage{xcolor}
\usepackage{diagbox}
\usepackage{siunitx}
\usepackage{caption}
\usepackage{pifont}
\usepackage{censor}
\newcommand{\cmark}{\ding{51}}%
\newcommand{\xmark}{\ding{55}}%
\usepackage{array}
\newcolumntype{P}[1]{>{\centering\arraybackslash}p{#1}}
\usepackage[caption=false]{subfig}
\def\BibTeX{{\rm B\kern-.05em{\sc i\kern-.025em b}\kern-.08em
    T\kern-.1667em\lower.7ex\hbox{E}\kern-.125emX}}
\begin{document}

\title{Calibration-free online test-time adaptation for electroencephalography motor imagery decoding\\
}

\author{\IEEEauthorblockN{1\textsuperscript{st} Martin Wimpff}
\IEEEauthorblockA{
\textit{University of Stuttgart}\\
Stuttgart, Germany \\
martin.wimpff@iss.uni-stuttgart.de}
\and
\IEEEauthorblockN{2\textsuperscript{nd} Mario Döbler}
\IEEEauthorblockA{
\textit{University of Stuttgart}\\
Stuttgart, Germany \\
mario.doebler@iss.uni-stuttgart.de}
\and
\IEEEauthorblockN{3\textsuperscript{rd} Bin Yang}
\IEEEauthorblockA{
\textit{University of Stuttgart}\\
Stuttgart, Germany \\
bin.yang@iss.uni-stuttgart.de}
}


\maketitle

\begin{abstract}

Providing a promising pathway to link the human brain with external devices, Brain-Computer Interfaces (BCIs) have seen notable advancements in decoding capabilities, primarily driven by increasingly sophisticated techniques, especially deep learning.
However, achieving high accuracy in real-world scenarios remains a challenge due to the distribution shift between sessions and subjects. In this paper we will explore the concept of online test-time adaptation (OTTA) to continuously adapt the model in an unsupervised fashion during inference time. Our approach guarantees the preservation of privacy by eliminating the requirement to access the source data during the adaptation process. 
Additionally, OTTA achieves calibration-free operation by not requiring any session- or subject-specific data.
We will investigate the task of electroencephalography (EEG) motor imagery decoding using a lightweight architecture together with different OTTA techniques like alignment, adaptive batch normalization, and entropy minimization.
We examine two datasets and three distinct data settings for a comprehensive analysis. 
Our adaptation methods produce state-of-the-art results, potentially instigating a shift in transfer learning for BCI decoding towards online adaptation.
\end{abstract}


\section{Introduction}
In the dynamic intersection of neuroscience and technology, researchers explore electroencephalography (EEG) motor imagery decoding as a promising avenue for connecting the human brain to external devices. This non-invasive and versatile approach enables the translation of mental simulations of movement into direct commands. It offers individuals with motor impairments, like those from spinal cord injuries or neurodegenerative diseases, a transformative pathway to restore functionality and improve quality of life through Brain-Computer Interfaces (BCIs).
In recent years, there has been a substantial increase in research focused on deep learning methods for BCIs \cite{roy2019deep, craik2019deep} as deep learning methods are able to implicitly extract intricate patterns from EEG signals to improve the decoding performance. The usability in real-world scenarios, however, remains limited as conditions change between development and deployment of the system \cite{wu2020transfer}.
It is therefore often necessary to record additional calibration data before deployment in order to adapt the model to the setting or subject which is both time and cost-intensive.
For BCIs, the most common change of conditions is a change of subject also referred to as the cross-subject scenario.
\begin{figure}[htp]
\includegraphics[width=\columnwidth]{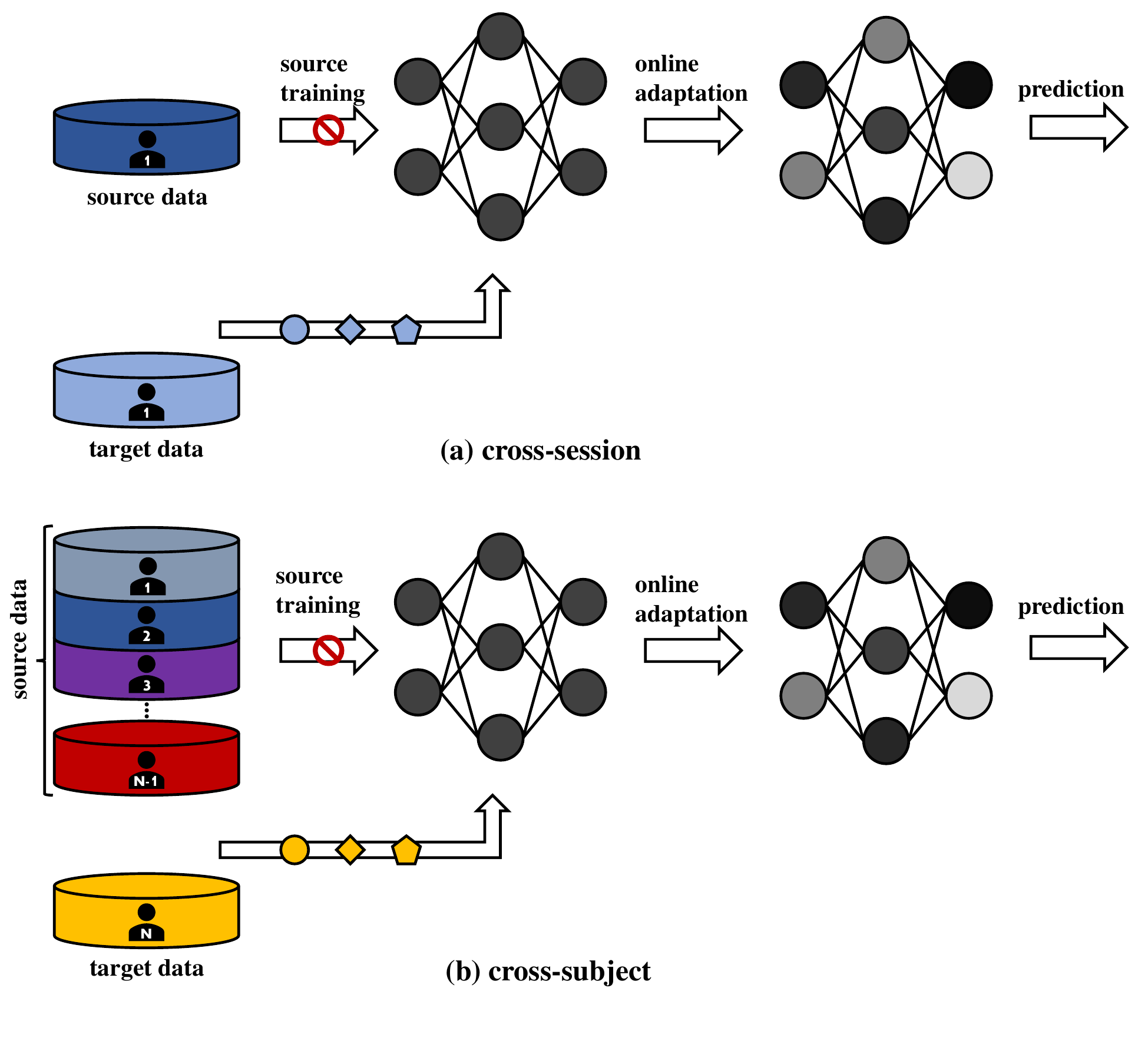}%
\captionsetup{belowskip=-20pt}
\caption{Methodology for the (a) cross-session and (b) cross-subject setting.}
\label{fig:settings}
\end{figure}
In this particular setting, data from multiple (source) subjects is used to train a model which is then tested on data from an unseen, sometimes called hold-out, (target) subject (see Fig. \ref{fig:settings}). 
Due to the high inter-subject differences, there is a large distribution shift and therefore a large drop in performance \cite{lawhern2018eegnet, kostas2020thinker}. While this drop in performance is particularly large in the cross-subject scenario, it is also present in the cross-session setting where train and test data are recorded from the same subject in different sessions. 
Therefore, the BCI community is actively developing solutions which minimize the amount of calibration data while maximizing the performance \cite{ko2021survey}. 
Most approaches focus on the common cross-subject scenario but there has also been work on cross-dataset \cite{xu2021enhancing} or even cross-task solutions \cite{aristimunha2023evaluating}.\newline
Approaches targeting the transfer of knowledge from a source domain or task to a distinct (but related) target domain or task fall under the umbrella of transfer learning.
As the task, the method, the learning scenario and the data properties vary between approaches, different methods can be broadly categorized in different sub-categories. Two well known major concepts are domain generalization (DG) and domain adaptation (DA).
The objective of DG is to train a model using one or more source domains, enabling it to generalize effectively to any unseen (related) target domain. DA on the other hand learns a model that generalizes to a specific, known target domain. The major advantage of DG is that the model is only trained once on the source domain and the target domain can be unknown during training. The limitation of DG is that it can't compensate large and especially unexpected distribution shifts between source and target domain as it does not leverage the target data.
The most common DG approaches for EEG are data augmentation methods \cite{ko2021survey, rommel2022data} and alignment strategies \cite{he2019transfer, kostas2020thinker, zanini2017transfer}. 
The basic idea behind data augmentation is to manipulate the existing data to create new artificial samples in order to increase the amount of data without recording new data to make the model more robust i.e., invariant to certain transformations. The major challenge with data augmentation is to find a transformation that manipulates the samples while retaining the spatio-spectral-temporal information. Finding such transformations is extremely difficult for EEG and therefore most simple transformations underperform the baseline \cite{ko2021survey}.
Data alignment ensures that the covariance matrix of each trial is aligned using a reference matrix. Importantly, both, the source and target data needs to be aligned using one reference matrix per domain. This makes data alignment a special case of DG, as the target domain is only arbitrary as long as the reference matrix is known at inference time.
In comparison to DG, DA does compensate distribution shifts by leveraging the target data at the cost of needing additional labeled or unlabeled target data for adaptation. The most popular approach in DA is to finetune a pre-trained model using target data. Importantly, such finetuning approaches require enough target data to avoid overfitting \cite{kostas2020thinker}. Additionally, many DA methods, especially unsupervised ones, need to access the data from the source and target domain simultaneously which is often not possible due to privacy regulations. \newline
Overcoming these limitations, the field of (source-free) online test-time adaptation (OTTA) emerged \cite{schneider2020improving, wang2020tent, liang2020we, wang2022continual, marsden2022gradual, dobler2023robust, marsden2023universal, yuan2023robust, liang2023ttasurvey}. OTTA distinguishes itself from DA in that no source data is required during adaptation. Further, the adaptation is done online during inference, by using unlabeled test-time samples only, eliminating the need for additional calibration data from the target domain \cite{liang2023ttasurvey}.\newline
As we are the first to apply OTTA to EEG motor imagery decoding, there are only very few approaches \cite{an2020few, duan2020ultra, xu2021enhancing, xia2022privacy, mao2023online, wang2023privacy} which are partially comparable to our work. \cite{an2020few} and \cite{duan2020ultra} are classic DA methods which employ few-shot learning i.e., they use a few labeled target samples together with labeled source data to adapt their model.\cite{xu2021enhancing} uses a source-free cross-dataset scenario and compensate the distribution shift with offline alignment and Adaptive Batch Normalization (AdaBN)\cite{li2016revisiting}.
\cite{xia2022privacy} and \cite{wang2023privacy} are offline, cross-subject, source-free unsupervised DA approaches for EEG motor imagery decoding and intracranial EEG epilepsy detection, respectively. \cite{mao2023online} is the only online source-free approach for EEG decoding. In contrast to our work, they use a very large dataset and perform seizure prediction instead of motor imagery decoding. The size of their datasets allows them to use a mean teacher \cite{wang2022continual, dobler2023robust} for the adaptation.
\newline
In this study, we will explore the applicability of source-free online test-time adaptation methods for EEG motor imagery decoding on two different datasets. We will examine three distinct scenarios: cross-session, cross-subject and the continual cross-subject setting. Our adaptation approaches demonstrate cutting-edge performance without requiring calibration data while preserving privacy. These advancements enhance the usability of BCIs and could potentially instigate a paradigm shift in transfer learning for BCI decoding, transitioning from offline adaptation to online adaptation.
The source code is available at https://github.com/martinwimpff/eeg-otta.
\section{Method}
In this section, we will first describe the task, datasets and settings of our approach. Then the different aspects of our final adaptation method will be explained.
\subsection{Task and Datasets}
The decoding task in this work is single-trial EEG motor imagery decoding. The specific settings we will investigate are described in the next session.
We use the BCI Competition IV 2a \cite{brunner2008bci} and 2b dataset \cite{leeb2008bci} to evaluate our methods. Both datasets consist of data from 9 subjects recorded at \qty{250}{\hertz}. The 2a dataset contains four different classes (feet, left hand, right hand, tongue) whereas the 2b dataset only contains two classes (left hand, right hand). The 2a dataset was acquired using 22 electrodes, the 2b dataset only uses three electrodes. The 2a dataset was recorded in two sessions on different days with 288 trials per session. The 2b dataset contains five sessions of which the first three (400 trials) form the training set and the last two (320 trials) form the test set. The only preprocessing we use is a \qty{40}{\hertz} lowpass filter.
\subsection{Settings}
We investigate three different settings with three different data splits. The first two are visualized in Fig. \ref{fig:settings}. For simplicity, we will describe the splits using the terms of the 2a dataset. The first session and second session from the 2a dataset correspond to the first three sessions and the last two sessions of the 2b dataset, respectively. In all three settings, the reported results refer to the second session.
\subsubsection{Cross-session setting}
In this setting we train one model per subject using the data from the first session for source training and the data from the second session for online adaptation and evaluation. This setting is also known as the within-subject setting and should have a relatively small distribution shift.
\subsubsection{Cross-subject setting}
In the cross-subject setting the first sessions of eight source subjects are used for source training. The second session of the remaining unseen hold-out subject is used for online adaptation and evaluation. In this case, we expect a remarkably larger distribution shift than in the first setting.
\subsubsection{Cross-subject continual setting}
The source training and evaluation of this setting is similar to the previous cross-subject setting. However, the adaptation is done on the first and second session of the target subject to investigate whether the continuous adaptation improves the performance.
\subsection{Model and Source training}
As a deep learning model we will use BaseNet \cite{wimpff2023eeg} which is a lightweight architecture combining the advantages of EEGNet \cite{lawhern2018eegnet} and ShallowNet \cite{schirrmeister2017deep} into a powerful and versatile architecture. As in \cite{wimpff2023eeg}, all models are trained for a fixed number of epochs (1000) using a learning rate scheduler with a linear warmup (20 epochs) and a cosine decay. Additionally, we investigate the use of label smoothing \cite{szegedy2016rethinking}.
We run each experiment with five different random seeds and report the average test accuracy and standard deviation of these five runs across all subjects.
\subsection{Input buffer}
One particular challenge for single-instance OTTA is that the model has to make a prediction immediately after receiving a new input sample. As we are using online alignment and online batch statistics it is necessary to keep previous target samples in an input buffer to reliably estimate the current reference matrix and batch statistics respectively. The size of this input buffer, i.e., how many samples to keep, is a trade-off between performance, memory, and how fast the distribution shift occurs. Even with large memory capacities one should keep in mind that the target data distribution changes over time and it therefore can be detrimental to keep a large input buffer. We use a buffer size of 32 samples in our experiments and update the buffer using the first in first out principle.
\subsection{Alignment}
Covariance alignment is a common transfer learning strategy in EEG decoding because of its simplicity and effectiveness \cite{he2019transfer, kostas2020thinker, zanini2017transfer}. The main assumption is that the difference between two sessions or subjects can be captured in a reference state which can then be used to reduce the distribution shift between sessions or subjects. The basic idea for alignment is to estimate a mean covariance matrix $\overline{R}$ per domain using the $b$ samples in the buffer. For euclidean alignment that means computing the arithmetic mean of all covariance matrices \cite{he2019transfer}:
\begin{equation}
    \overline{R} = \frac{1}{b}\sum_{i=1}^{b}\gamma_i\cdot X_iX_i^T, \quad \gamma_i=1\forall i, \quad X_i\in\mathbb{R}^{C\times T}
    \label{eq:ea}
\end{equation}
For riemannian alignment (RA) \cite{zanini2017transfer}, the riemannian or geometric mean of all covariance matrices is computed instead.
The current trial $X_b$ is then aligned by:
\begin{equation}
    \Tilde{X}_b=\overline{R}^{-1/2}X_b
\end{equation}
Equation \ref{eq:ea} can additionally be manipulated using a linear or exponential weighting factor $\gamma_i$. Both approaches weight current trials stronger than older samples in the buffer. We use exponential weighting with a momentum of $0.1$ in our experiments.
\subsection{Batch Normalization}
Batch normalization (BN) was originally developed to reduce the internal covariate shift due to the change in network parameters during training \cite{ioffe2015batch}. During training, BN layers normalize each feature channel using the statistics of the current batch:
\begin{equation}
    y=\frac{x-\mathrm{E}[x]}{\sqrt{\mathrm{Var}[x]+\epsilon}}\cdot\gamma+\beta, \quad \epsilon=10^{-5}
\end{equation}
$\gamma$ and $\beta$ are learnable affine parameters.
During training, each batch normalization layer keeps a global estimate for the mean $\mu$ and variance $\sigma^2$ of the whole training set which is updated using an exponential moving average. 
These global training statistics are then applied to normalize the test data during inference. This makes the model decisions deterministic but might also cause troubles under distribution shifts \cite{schneider2020improving}.
\cite{li2016revisiting} developed AdaBN which simply replaces the global training statistics by global statistics of the test set to reduce the distribution shift. However, those global statistics are estimated offline.
\cite{schneider2020improving} proposed an online version of AdaBN which only uses the available samples from the test data. In our experiments we will use a similar approach where we estimate the statistics using all samples in the buffer. 
If the number of samples in the buffer is very small, it can be useful to employ the source statistics $\mu_s, \sigma^2_s$ as a prior \cite{schneider2020improving, marsden2022gradual}:
\begin{align}
   \mu_{i} &= (1-\alpha)\mu_{s}+\alpha\mathrm{E}[x]\\
   \sigma^2_{i} &= (1-\alpha)\sigma^2_{s}+\alpha\mathrm{Var}[x]
\end{align}
where $x$ represents the test-time samples in the buffer.
Another option is to use an exponential moving average \cite{yuan2023robust} of the target statistics which are initialized with the source statistics. We will refer to these variations as BN-$\alpha$ and BN-EMA. Consequently, BN-1 denotes that only the target statistics are used.
\subsection{Entropy Minimization}
So far, the model parameters $\theta$ of our deep learning model $f_\theta$ remained untouched during the adaptation. However, keeping the parameters learned on the source domain is not ideal. Therefore many approaches use an unsupervised loss to adapt the network parameters to the target domain. In this work we will minimize the entropy $H(\hat{y})=-\sum_cp(\hat{y}_c)\log p(\hat{y}_c)$ of the model predictions $\hat{y}=f_\theta(x)$. Minimizing the entropy drives the model to make more confident predictions during test-time without requiring labeled target data. As we do single-instance OTTA, the loss is only calculated and backpropagated every $b$ samples, i.e., whenever the buffer is completely renewed. We use the Adam optimizer with default parameters and a learning rate of $5\cdot10^{-4}$ for adaptation.\newline
Since the entropy objective and therefore the entire adaptation process relies on the model's initial confidence, we explore the connection between label smoothing in source training and entropy minimization during adaptation. 
\begin{table}[h!]
\centering
\captionsetup{belowskip=-10pt}
\caption{Results for the cross-session setting}
\label{tab:cross-sess1}
\begin{tabular}{ |p{2.9cm}|P{0.4cm}||P{1.95cm}|P{1.95cm}|  }
 \hline
 method & EM & BCIC IV 2a ($\%$) &BCIC IV 2b ($\%$)\\
 \hline
 source  &\xmark & $76.16\pm 0.45$ &$84.75\pm 0.24$\\
 \hline
 EA   &\xmark& $75.43\pm 0.35$ &$81.25\pm 0.93$\\
 EA(linear)  &\xmark & $75.32\pm 0.39$ &$81.15\pm 0.89$\\
 EA(EMA)  &\xmark & $75.32\pm 0.38$ &$81.12\pm 0.96$\\
 RA &\xmark  & $76.91\pm 0.46$ &$85.32\pm 0.52$\\
 RA(linear) &\xmark  & $76.81\pm 0.42$ &$85.30\pm 0.56$\\
 RA(EMA) &\xmark  & $76.77\pm 0.39$ &$85.37\pm 0.29$\\
 \hline
 BN-1 &\xmark  & $77.72\pm 0.69$ &$85.54\pm 0.34$\\
 BN-0.5&\xmark & $77.86\pm 0.47$ &$85.84\pm 0.31$ \\
 BN-EMA &\xmark  & $77.81\pm 0.49$ &$85.50\pm 0.43$\\
 \hline
 RA(linear)-BN-1  &\xmark & $78.16\pm 0.70$ &$85.47\pm 0.20$\\
 RA(EMA)-BN-1 &\xmark  & $78.29\pm 0.61$ &$85.47\pm 0.20$\\
 RA(linear)-BN-0.5 &\xmark  & $78.19\pm 0.62$ &$85.76\pm 0.41$\\
 RA(EMA)-BN-0.5 &\xmark  & $78.33\pm 0.72$ &$85.79\pm 0.40$\\
 RA(linear)-BN-EMA &\xmark  & $78.09\pm 0.56$ &$85.55\pm 0.32$\\
 RA(EMA)-BN-EMA &\xmark  & $78.14\pm 0.53$ &$85.54\pm 0.38$\\
 \hline
 RA(EMA)-BN-1($\delta$=0) &\cmark & $78.22\pm0.54$ & $86.11 \pm 0.41$\\
 \textbf{RA(EMA)-BN-1(}$\boldsymbol{\delta}$\textbf{=0.5)} &\cmark & $\boldsymbol{79.74\pm0.57}$ & $\boldsymbol{86.59\pm0.34}$\\
 \hline
\end{tabular}
\end{table}
Label smoothing \cite{szegedy2016rethinking} simply replaces the original hard class label $y_c\in\{0, 1\}$ by $y_c^{\text{LS}}=y_c(1-\delta)+\delta/C$ with $C$ being the total number of classes and $0<\delta<1$. 
Without label smoothing, the model tends to learn sharp decision boundaries and produces over-confident predictions. This leads to high confidences and low entropy around the decision boundary. It is difficult to optimize such boundaries using entropy minimization, as the model already has a low entropy for most of the samples. With label smoothing the model creates smoother boundaries that can be easier optimized during adaptation.

\section{Results}
This section is structured according to the different settings introduced in the previous section and compares our OTTA approaches against a source model without adaptation.
\subsection{Cross-session}
Table \ref{tab:cross-sess1} shows the test accuracies for the 2a and 2b dataset in the cross-session setting. The second column indicates the use of entropy minimization (EM). Interestingly, euclidean alignment (EA) yields results below the source performance for both datasets. Riemannian alignment (RA) on the other hand results in test accuracies slightly above source performance. For the 2a dataset, the adaptive BN approaches further improve the results. Combining both methods leads to the best results which are $\sim2\%$ and $\sim1\%$ above the source accuracy for the 2a and 2b dataset respectively.

Our method employing RA(EMA) together with entropy minimization to adapt the model parameters further improves the results to $79.74\pm0.57\%$ ($\delta=0.5$) and $86.59\pm0.34\%$ ($\delta=0.5$). Without label smoothing ($\delta=0$) the results are considerably lower ($78.22\pm0.54\%$ and $86.11\pm0.41\%$).

\subsection{Cross-subject}
The results for the cross-subject setting are shown in Table \ref{tab:cross-sub1}. 
Using online alignment leads to a $4-5\%$ increase in performance for the 2a dataset, with RA consistently outperforming EA. For the 2b dataset the improvement is only around $1\%$ which might be due to the low number of sensors. Applying the target statistics instead of the source statistics during batch normalization outperforms all alignment methods for both datasets. 
\begin{table}[h!]
\centering
\captionsetup{belowskip=-10pt}
\caption{Results for the cross-subject setting}
\label{tab:cross-sub1}
\begin{tabular}{ |p{2.9cm}|P{0.4cm}||P{1.95cm}|P{1.95cm}|  }
 \hline
 method & EM & BCIC IV 2a ($\%$) &BCIC IV 2b ($\%$)\\
 \hline
 source  &\xmark & $57.38\pm 0,92$ &$78,75\pm 0,95$\\
 \hline
 EA &\xmark  & $62.48\pm 0.68$ &$79.21\pm 0.25$\\
 EA(linear) &\xmark  & $62.74\pm 0.65$ &$79.19\pm 0.22$\\
 EA(EMA)  &\xmark & $62.45\pm 0.69$ &$79.15\pm 0.18$\\
 RA &\xmark  & $63.64\pm 0.41$ &$79.96\pm 0.29$\\
 RA(linear) &\xmark  & $63.86\pm 0.38$ &$80.11\pm 0.23$\\
 RA(EMA) &\xmark  & $63.68\pm 0.29$ &$80.22\pm 0.24$\\
 \hline
 BN-1 &\xmark  & $64.32\pm 0.91$ &$81.70\pm 0.26$\\
 BN-0.5&\xmark & $64.18\pm 0.85$ &$81.88\pm 0.16$ \\
 BN-EMA &\xmark  & $64.32\pm 0.97$ &$81.78\pm 0.25$\\
 \hline
 RA(linear)-BN-1 &\xmark  & $65.16\pm 0.59$ &$81.95\pm 0.46$\\
 RA(EMA)-BN-1 &\xmark  & $65.40\pm 0.54$ &$81.97\pm 0.47$\\
 RA(linear)-BN-0.5  &\xmark & $65.59\pm 0.71$ &$82.18\pm 0.51$\\
 RA(EMA)-BN-0.5 &\xmark  & $65.58\pm 0.64$ &$82.22\pm 0.54$\\
 RA(linear)-BN-EMA  &\xmark & $65.29\pm 0.48$ &$81.99\pm 0.39$\\
 RA(EMA)-BN-EMA  &\xmark & $65.09\pm 0.45$ &$81.96\pm 0.57$\\
 \hline
 RA(EMA)-BN-1($\delta$=0) &\cmark & $65.31\pm0.57$ & $82.77 \pm 0.61$\\
 \textbf{RA(EMA)-BN-1(}$\boldsymbol{\delta}$\textbf{=0.4)} &\cmark & $\boldsymbol{67.31\pm1.38}$ & $\boldsymbol{83.47\pm0.36}$\\
  \hline
\end{tabular}
\end{table}
Consequently, combining both approaches leads to the highest results which are $\sim8\%$ and $\sim3\%$ above the source performance for the 2a and 2b dataset respectively.

\begin{figure}
    \centering
    \includegraphics[width=\columnwidth]{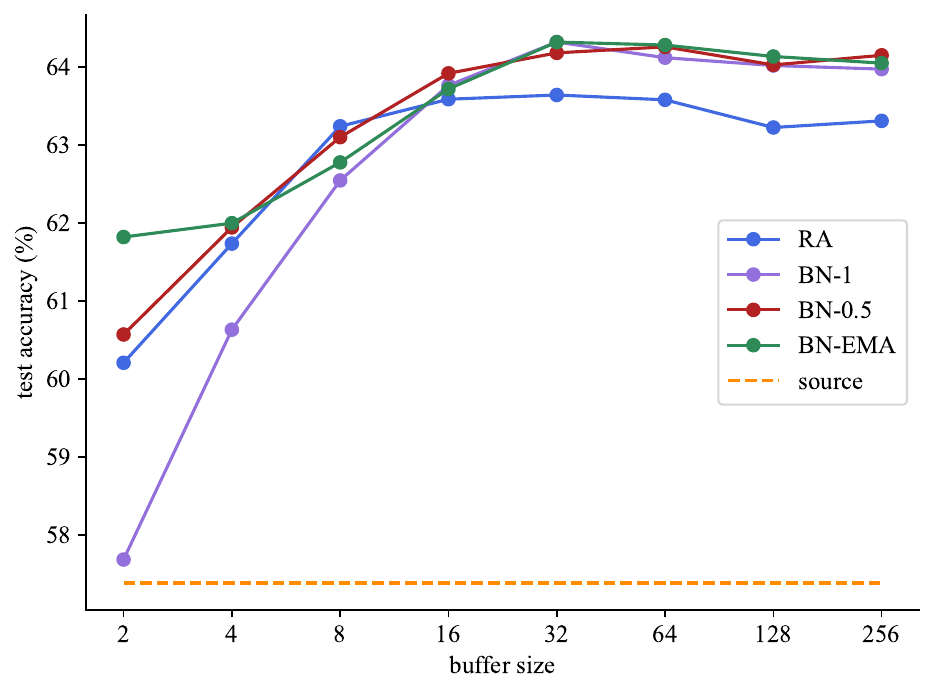}
    \captionsetup{belowskip=-20pt}
    \caption{Test accuracy and buffer size.}
    \label{fig:bn-buffer}
\end{figure}

\textit{Buffer size} As BN adaptation plays a significant role in mitigating potential domain shifts, we investigate different buffer sizes in Fig. \ref{fig:bn-buffer}. As in our experiments in Table \ref{tab:cross-sub1}, for a buffer size of 32, the difference between the BN approaches is neglectable. The same holds for larger batch sizes. 
Therefore, we employ BN-1 for our method as it also resembles the common setting during training.
However, as BN-1 only relies on the test statistics, small buffer sizes are not sufficient for a reliable estimation of the BN statistics. In this case, exploiting the source statistics, as in BN-0.5, or using an exponential moving average, as in BN-EMA, is beneficial. 

\begin{figure}
    \centering
    \includegraphics[width=\columnwidth]{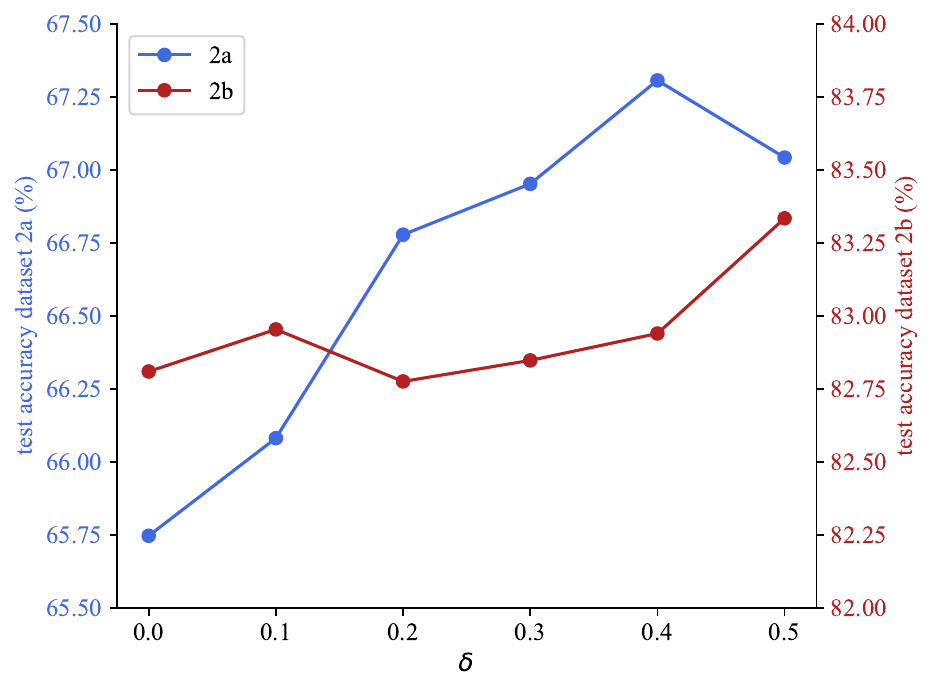}
    \captionsetup{belowskip=-20pt}
    \caption{Label smoothing and entropy minimization.}
    \label{fig:tent-ls-both}
\end{figure}

\textit{Entropy minimization} In addition, we also investigate the impact of the label smoothing hyperparameter $\delta$ during source training together with entropy minimization. Fig. \ref{fig:tent-ls-both} displays the results for both datasets with additional RA(EMA). For the 2a dataset there is a clearly visible trend that label smoothing helps the optimization during OTTA. The best accuracy ($67.31\pm 1.38\%$) can be obtained for a label smoothing factor $\delta=0.4$ which is significantly above the runs without label smoothing ($65.75\pm0.57\%$). As expected, the differences are lower for the 2b dataset. The best result ($83.47\pm0.36$) is only slightly better than the adaptation without label smoothing ($82.77\pm0.61$). For both datasets, the results are $1-2\%$ above the results of their respective counterpart (RA(EMA)-BN-1) without entropy minimization.

\subsection{Continual setting}
The continual setting leads to almost a two percent increase in performance for the 2a dataset ($67.31\pm 1.38\%\rightarrow69.01\pm0.96\%$) and a slight decrease in variance between the runs. For the 2b dataset our first experiments yielded results ($82.12\pm0.73$) below the results in the cross-subject setting. We believe that this is due to the fact that the training set contains three sessions, of which two are recorded without feedback (offline) and one with feedback (online). The test set contains two sessions with feedback\cite{leeb2008bci}. Therefore we ran two additional experiments where we used either the first two or only the third session for the first adaptation phase. That resulted in test accuracies of $81.97\pm0.4\%$ and $83.47\pm0.36\%$ respectively. This shows that the continual setting can be a powerful tool to exploit additional data as long as the recording settings between the different settings are similar enough.

\begin{figure}
    \centering
    \includegraphics[width=\columnwidth]{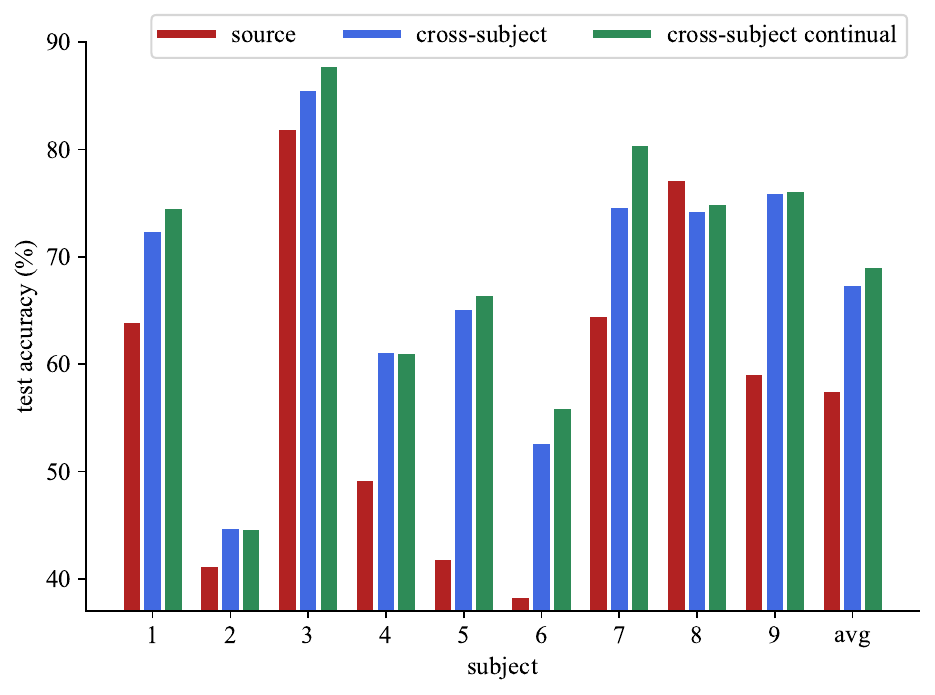}
    \captionsetup{belowskip=-20pt}
    \caption{Results per subject for the 2a dataset.}
    \label{fig:2a-per_subject}
\end{figure}

Fig. \ref{fig:2a-per_subject} shows the results per subject for the 2a dataset. As with all BCI experiments, the differences between the various subjects are very high. $70\%$ accuracy is often considered as the threshold after which a BCI is usable. With this threshold, our method enables 3 out of 7 subjects who were previously below the threshold to use BCIs.

\section{Discussion}
\textit{Related work}
To validate our results, we tried to compare them against the literature. However, as we are the first to apply OTTA to EEG motor imagery decoding, such comparisons have only limited explanatory power. 
The only setting comparable to our approach (cross-subject continual setting) is the privacy-preserving offline unsupervised DA setting \cite{xia2022privacy}. Their final test accuracy on the 2a dataset is $57.35\%$ which is at the level of our source performance.
The two few-shot approaches \cite{duan2020ultra, an2020few} are difficult to compare to our setting as they use labeled target data. The zero-shot performance (comparable to our cross-subject setting) of \cite{duan2020ultra} is at the level of our source model for the 2a dataset ($\sim 10\%$ below our results with adaptation). With labeled target data, their performance is higher ($76.3\%$). \cite{an2020few} used the 2b dataset and reported results ($74.6\%$) below our source model despite using labeled target data for adaptation. 
\cite{xu2021enhancing} operates in a cross-dataset setting and report a test accuracy of $77.8\%$ for the 2a dataset using only the left and right hand trials. 
This short attempt to compare our approach to the literature highlights how difficult it is to benchmark our approach due to the new setting.
Therefore we chose to compare our approach against a strong baseline model \cite{wimpff2023eeg} which is trained on the source data. The improvements relative to this baseline can then be directly attributed to the specific adaptation method because the model, setting, hyperparameters, etc. remain constant.

\textit{Alignment}
Using online alignment improved the performance of the model, especially for the cross-subject setting. RA consistently outperformed EA which might be due to the fact that the geometric mean is generally considered to be more robust than arithmetic mean. The minor improvement for the cross-session setting can probably be attributed to the small change of reference states between the sessions.

\textit{Batch Normalization} Replacing the source statistics in the BN layers with target statistics outperformed all alignment approaches across both datasets. This simple measure alone leads to a $\sim7\%$ and $\sim3\%$ performance increase in the cross-subject setting for the 2a and 2b dataset respectively. This demonstrates the significance of the distribution shift as well as the effectiveness of very simple OTTA methods for EEG motor imagery. The combination of online alignment and adaptive BN further improved the results. 

\textit{Entropy minimization}
With entropy minimization, the test accuracy was improved by $\sim3.5\%$ and $\sim2\%$ in the cross-session setting for the 2a and 2b dataset respectively. For the cross-subject setting the results are $\sim10\%$ and $\sim5\%$ above the baseline. 
Incorporating label smoothing during source training emerged as essential for the optimization during adaptation.
This technique prevents the source model from generating overly confident predictions and creating sharp decision boundaries, thereby preserving a smoother loss surface for optimization.

\textit{Continual adaptation}
To investigate the use of our method for situations where a model is continuously adapted for one subject over multiple sessions we ran experiments using both sessions successively. For the 2a dataset this improved the results by almost two percent. For the 2b dataset, however, this setting was more challenging as the training set consists of offline and online sessions. The continuous setting worked for the 2b dataset if only the online session (of the training set) was used for adaptation. Using all three sessions or both offline sessions resulted in lower performances. This suggests that the continuous setting is only advantageous if the settings between the sessions are similar enough. Otherwise the model should be adapted to every sessions independently.
\section{Conclusion}
In this work we explored the use of online test-time adaptation for EEG motor imagery for three different settings on two distinct datasets. We achieved state of the art results, improving existing methods by a large margin (up to $12\%$) while preserving the privacy of the source subjects. Our method can be used cross-session within the same subject or cross-subject, hence greatly improving the applicability of BCIs in real-world scenarios where unknown distribution shifts occur during deployment of the system.
\section*{Acknowledgment}
This paper is a result of the BMBF-funded Quantum Human Machine Interfaces (QHMI) project under the QSens - Quantum Sensors of the Future cluster.

\end{document}